\documentclass[aps,prl,twocolumn,showpacs,a4paper]{revtex4}
\usepackage{bm,graphicx,amsmath,color,braket}

\begin{document}
\title{Pre-entaglement shows that individual quantum systems do not possess states}

\author{D. J. Miller}
\email[]{d.miller@sydney.edu.au}
\author{Matt Farr}

\affiliation{Centre for Time, University of Sydney NSW 2006, Australia}
\date{\today}

\begin{abstract}
We consider an experiment in which quantum tomography data is collected before and after an entangling measurement performed on two independently prepared, maximally-mixed ensembles. We show that each sub-ensemble that is, as expected, entangled after the measurement is also entangled before it. We call the latter pre-entanglement. If individual systems possessed quantum states, their independent preparations would mean the sub-ensembles were in separable states prior to the entanglement event. The contradiction can be resolved by assigning states only to ensembles and not individual quantum systems. Pre-entanglement found here at the quantum level makes very doubtful the assumption of preparation independence at the ontic level required for a recent theorem.
\end{abstract}

\maketitle

There is a general category of interpretations of quantum mechanics, associated in one way or another with Einstein \cite{einstein}, Pearle \cite{pearle}, Park \cite{park}, Ballentine \cite{ballentine,ballentine2} and others \cite{home}, in which a quantum state $\psi$ is a property of an ensemble and individual systems are not assigned quantum states. We will refer to the above category of interpretation as the ensemble-system interpretation (ESI). The more orthodox interpretation is that individual quantum systems may be assigned states; we refer to this as the quantum-system interpretation (QSI). It follows that according to the ESI, there are no proper mixtures in quantum mechanics. In a \textit{proper mixture} \cite{espagnat} one knows (perhaps from the way the ensemble was prepared) that a certain set of states $\psi_i$ appears with probability $p_i$ among the constituents as a whole (although one may be ignorant of which constituent of the ensemble has which state).  In an \textit{improper mixture}, the constituents cannot be assigned their own states on any grounds. The most familiar example of improper mixtures are each of the subsystems in a set of singlet states. Our aim is to show that the assumption that there are proper mixtures of quantum states leads to a contradiction and thereby prove that (some version of) ESI is the only viable interpretation of quantum mechanics.

To demonstrate our result, we consider ensembles of qubits (the simplest quantum system is sufficient) and assume (it will turn out wrongly) a quantum state $\psi$ can be identified with each qubit that is prepared. Then if two mixtures of qubits are paired off, each pair is in a product state and any ensemble (or sub-ensemble) formed from them must be separable (a separable state is defined to be one which is not entangled). The contradiction arises because we show that one can identify and measure a sub-ensemble in an \textit{entangled} state within the apparently separable ensemble formed by pairing-off two apparently proper ensembles of maximally-mixed qubits.

There are two reasons that hint that this may be possible. Firstly, it is known from delayed-choice entanglement swapping (D-CES) experiments \cite{peres2,ma,megedish} that one can identify entangled sub-ensembles formed from pairing two improper ensembles of qubits (i.e.\ qubits whose state is the reduced density operator of an entangled state). Since there is no discretionary experimental control of the preparation, it does not decide anything directly about the ESI versus QSI question. Secondly, it is possible that an ensemble composed of a proper mixture of entangled pairs is in a separable state overall. Timpson and Brown \cite{timpson} have coined the term \textit{improper separability} for those cases. 

Specifically, we show that one or more of the following statements is incorrect. The statements are sufficient for our purposes but some of them are not necessary and could be relaxed.

[1.]\ \textit{Preparation of a pure state.} A quantum system can be prepared in a pure state by measuring an observable with non-degenerate eigenvalues $o_i$ and corresponding eigenstates $|\psi_i  \rangle$. When $o_i$ is recorded by a classical system as being the result of the measurement, the qubit has been prepared in the state $|\psi_i  \rangle$.

[2.]\ \textit{Persistence of a quantum state.} A quantum system prepared in a known state $|\psi_i \rangle$ continues to be in that state no matter which ensemble the qubit is assigned to (if $\hat{U}$ is the time-evolution operator over the relevant period we are, of course, assuming $\hat{U}|\psi_i \rangle=|\psi_i \rangle$ for all $i$).

[3.]\ \textit{Ensemble formation.} Alice can prepare the states $|\psi_{m} \rangle$, $m=1, \ldots M$ as in [1.]\ and keep a record of the states that are prepared by listing the run number $i$ and the eigenvalue for each preparation. If Alice prepares a series of $N$ qubits in states $|\psi_{m_1},  \rangle, |\psi_{m_2},  \rangle, \ldots, |\psi_{m_i} \rangle, \ldots$, $i=1,N$, the series forms an ensemble of pure states with the ensemble state
\begin{equation}
\hat{\sigma}_A =  \frac{1}{N}\sum_{i=1}^N |\psi_{m_i} \rangle \langle \psi_{m_i} |.
\end{equation}

Bob can prepare the states $|\phi_{m} \rangle$, $m=1,M$ as in [1.]\ and keep a record similar to Alice's. Let us assume Bob prepares the series $|\phi_{m_1},  \rangle, |\phi_{m_2},  \rangle, \ldots, |\phi_{m_i} \rangle, \ldots$, $i=1,N$ with the ensemble state 
\begin{equation}
\hat{\sigma}_B  =  \frac{1}{N}\sum_{i=1}^N |\phi_{m_i} \rangle \langle \phi_{m_i} | .
\end{equation}
Then one can form an ensemble of pairs of qubits with the state
\begin{equation}
\hat{\rho}_{AB}  =  \frac{1}{N^2}\sum_{i=1}^N |\psi_{m_i} \rangle \langle \psi_{m_i} | \otimes |\phi_{m_i} \rangle \langle \phi_{m_i} | . 
\end{equation}

The ensemble is a proper mixture because its members are identifiable by the records of run numbers and eigenvalues. It is separable because the partial transpose of $\hat{\rho}$ is the state itself and therefore has positive eigenvalues \cite{peres3}. An ensemble formed in the above way will be called a proper, separable ensemble.

[4.]\ \textit{Sub-ensemble formation.} (a) By choosing a subset of runs of an experiment $i \in \{\eta\}$ based on experimentally reproducible criteria, one can form a systematically-selected sub-ensemble.

(b) If this procedure is applied to the labelled qubits that Alice and Bob have prepared, the state of the sub-ensemble is
\begin{equation}
\hat{\rho}_{\{\eta\}}  = \frac{1}{N^2_{\eta}}\sum_{i \in \{\eta\}} |\psi_{m_i} \rangle \langle \psi_{m_i} | \otimes |\phi_{m_i} \rangle \langle \phi_{m_i} | 
\end{equation}
where $N_{\eta}$ is the number of members of the set $\{\eta\}$. All sub-ensembles formed in that way from Alice's and Bob's qubits are proper, separable ensembles. 

[5.]\ \textit{Quantum tomography.} The state of a sub-ensemble of pairs of qubits with $i \in \{\eta\}$ can be determined by quantum tomography (QT) \cite{altepeter}. The quantities needed for QT can be calculated from the procedures of standard quantum mechanics.

In [4.]\ the requirement ``experimentally reproducible criteria" for a ``systematically-selected sub-ensemble" means that if the whole experiment is repeated many times, one can identify a sub-ensemble in the same state in the same way each time. A randomly chosen sub-ensemble may \textit{appear} to possess its own ``state" but if that differs from the ensemble from which it is chosen, it is due merely to a statistical fluctuation entirely consistent with the sub-ensemble having the same state as the ensemble as a whole.

As shown in Fig.~1, in each run $i$ of the experiment, Alice chooses at random but with equal probability to prepare a qubit in one of the orthogonal states $\ket{\psi_1}$ or $\ket{\psi_2}$. Bob does the same for the orthogonal states $\ket{\phi_1}$ or $\ket{\phi_2}$ independently of (e.g.\ at space-like separation from) Alice. The states of Alice's and Bob's ensembles are respectively
\begin{equation}
\hat{\sigma}_A = \hat{\sigma}_B =\hat{I}/2
\end{equation}
where $\hat{I}$ is the identity operator in a two-dimensional Hilbert space. Alice at $P_A$ and Bob at $P_B$ subject each of their respective qubits to one projective measurement suitable for quantum tomography (QT) \cite{altepeter} chosen independently and randomly. They each record the run number $i$ and the result and pass their qubit to the next stage irrespective of the outcome. Next, an entangling measurement of the observable $Q$ with the entangled eigenstates $\ket{\Phi_n}, n=1,4$ (e.g.\ the Bell states \cite{nc}) is performed on each pair $i$. Finally, independent QT measurements $R_C$ and $R_D$ are performed on each qubit emerging from the $Q$ measurement and the run number and result is recorded.

\begin{figure}
\includegraphics[scale=0.4]{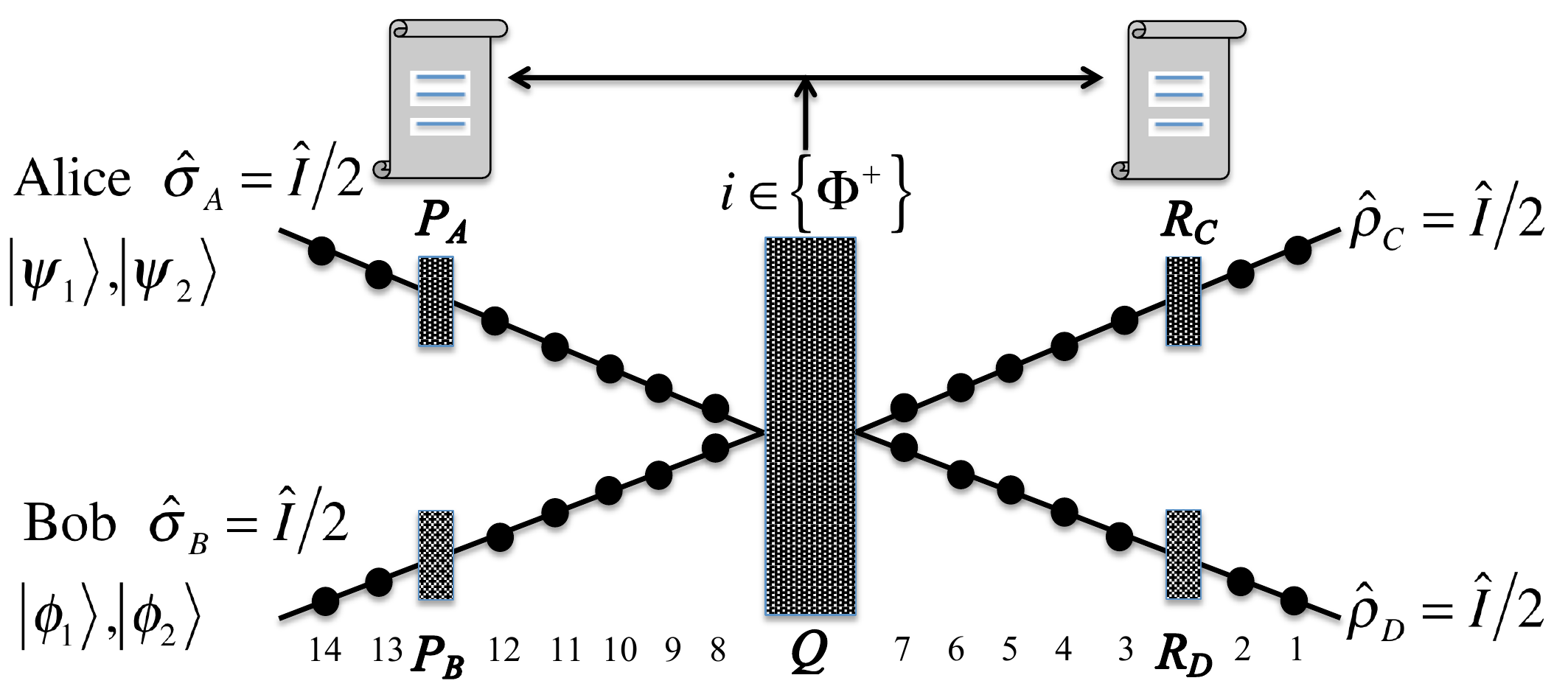}
\caption{Runs $i=1$ to $i=14$ of a much longer series of runs of the experiment are shown. In each run, Alice chooses at random but with equal probability to prepare qubits in one of the orthogonal states $\ket{\psi_1}$ or $\ket{\psi_2}$. Bob does the same for the orthogonal states $\ket{\phi_1}$ or $\ket{\phi_2}$. Each qubit is subject to a quantum tomography measurement $P_A$ and $P_B$ respectively and then the qubits are entangled at $Q$. After the entangling measurement $Q$, each emerging qubit is subject to another quantum tomography measurement $R_C$ or $R_D$ respectively. The runs $i \in \{ \Phi^+ \} $ for which $\ket{\Phi^+}$ is obtained at $Q$ are noted and a state is constructed from the $P_A$ and $P_B$ tomography data, and also from the $R_C$ and $R_D$ tomography data, for those runs. The state constructed from $P_A$ and $P_B$ prior to the entangling measurement at $Q$ is the same as the state constructed from $R_C$ and $R_D$ after $Q$.}
\end{figure}

For convenience, we will consider the runs $i \in \{\Phi^+\}$ for which the $Q$ measurement outcome is the Bell state $\ket{\Phi^+}  =  \frac{1}{\sqrt{2}}(\ket{0}_C \ket{0}_D+ \ket{1}_C \ket{1}_D)$. The runs $i \in \{\Phi^+\}$ form a systematically-selected sub-ensemble of qubits prepared by Alice and Bob which according to steps [1.] to [4.] has the state
\begin{equation}
\hat{\rho}_{\{\Phi^+\}}  =  \frac{1}{N^2_{\Phi^+}} \sum_{i \in \{\eta\}} |\psi_{m_i} \rangle \langle \psi_{m_i} | \otimes |\phi_{m_i} \rangle \langle \phi_{m_i} | \label{phirho}
\end{equation}
which is a proper, separable ensemble.

The state of a sub-ensemble of pairs can be constructed \cite{altepeter} from the experimental QT data recorded at the measurements $P_A$ and $P_B$ or from the data at $R_C$ and $R_D$. The data to construct the state of a sub-ensemble of pairs of runs $i \in \{E\}$ consists of joint probabilities like $\text{P}_{E}[j_A^{\pm} \& k_B^{\pm}]$ where $\text{P}_{E}[i_A^{\pm}]$ is the probability that the outcome of the measurement on Alice's qubit of an observable $\hat{\sigma}_i$ is the eigenvalue $i^{\pm}$, where $\hat{\sigma}_i \ket{i^{\pm}} = \pm \ket{i^{\pm}}$. If $\hat{\tau}_E$ is the state of the sub-ensemble
\begin{equation}
\text{P}_E[j_A^{\pm} \& k_B^{\pm}] = \text{Tr}(\hat{\tau}_E (\hat{P}_{j_A^{\pm}} \otimes  \hat{P}_{k_B^{\pm}})) \label{qt}
\end{equation}
where $\hat{P}_{i_A^{\pm}} = \ket{i^{\pm}}{_A}{_A}\bra{i^{\pm}}$. The unknown state $\hat{\tau}_E$ can be constructed from a full set of QT measurements of the probabilities $\text{P}_E[j_A^{\pm} \& k_B^{\pm}]$.

In the absence of experimental data, we can calculate the required results by applying standard quantum mechanics and hence anticipate the data that would be obtained in an experiment. This was the procedure first applied \cite{peres2} in D-CES and later confirmed in the laboratory (e.g.\ Ref. [\onlinecite{ma,megedish}]). Because the initial states are maximally mixed, the state for the whole ensemble continues to be $\hat{I}/2 \otimes \hat{I}/2$ after each stage of the experiment ($P_A/P_B$, $Q$, $R_C/R_D$). For the $R_C/R_D$ data, the required terms come from the probabilities conditional on $\Phi^+$ which are straightforward
\begin{subequations}
\begin{align}
\text{P}_{\Phi^+}[j_C^{\pm} \& k_D^{\pm}] & = \text{P}[j_C^{\pm} \& k_D^{\pm}|\Phi^+]  \\
& = \text{Tr}(\hat{P}_{\Phi^+} (\hat{P}_{j_C^{\pm}} \otimes  \hat{P}_{k_D^{\pm}})) . \label{rqt}
\end{align}
\end{subequations}
Comparing Eqs.~(\ref{qt}) and (\ref{rqt}), we see, entirely as expected, that the calculated QT data show that the state at the R measurements for the pairs of preparations $i \in \{\Phi^+\}$ is indeed $\ket{\Phi^+}$. 

We can follow the same procedure for the $P_A$ and $P_B$ measurements for the same ensemble of runs $i \in \{\Phi^+\}$
\begin{subequations}
\begin{align}
\text{P}_{\Phi^+}&[j_A^{\pm} \& k_B^{\pm}]  = \text{P}[j_A^{\pm} \& k_B^{\pm}|\Phi^+]  = \frac{\text{P}[j_A^{\pm} \& k_B^{\pm} \& \Phi^+]}{\text{P}[\Phi^+]} \\
& = \frac{\text{Tr}((\hat{P}_{j_A^{\pm}} \otimes  \hat{P}_{k_B^{\pm}})(\hat{I}/2 \otimes \hat{I}/2)(\hat{P}_{j_A^{\pm}} \otimes  \hat{P}_{k_B^{\pm}})\hat{P}_{\Phi^+})}{\text{Tr}((\hat{I}/2 \otimes \hat{I}/2)\hat{P}_{\Phi^+})}  \\
& = \text{Tr}(\hat{P}_{\Phi^+} (\hat{P}_{j_A^{\pm}} \otimes  \hat{P}_{k_B^{\pm}}) ) . \label{pqt}
\end{align}
\end{subequations}
The expression is simple because the source is unbiased \cite{pegg}. Comparing Eqs.~(\ref{qt}) and (\ref{pqt}), we see the result is exactly the same: the QT data for the ensemble can be calculated from the state $\ket{\Phi^+}$. Consequently, the sub-ensemble made up of the pairs $i \in \{\Phi^+\}$ is in an entangled state at both the stage of the $R$ measurements (Eq.~(8b), \textit{after} the entangling measurement and which we call post-entanglement, and at the stage of the $P$ measurements (Eq.~(9c), \textit{before} the entangling measurement which we call pre-entanglement. Post-entanglement is what is currently called entanglement. Pre-entanglement would show all the characteristics, like violation of the Bell inequality, nonlocality and quantum steering, etc.\ as post-entanglement. 

The same pre-entanglement state $\ket{\Phi^+}$ results from a calculation for the new procedure of direct measurement of quantum states \cite{kocsis,lundeen} and for weak measurement tomography \cite{hofmann} applied to the present scenario. Furthermore, what we have called pre-entanglement is used as a resource in measurement-device-independent quantum key distribution \cite{biham,rubenok,liu}.

The pre-entangled state $\ket{\Phi^+}$ of the sub-ensemble $i \in \{\Phi^+\}$ contradicts the state given in Eq.\ (\ref{phirho}) which, as noted above, is a separable state. This means one or more of the steps [1.]\ to [5.]\ in deriving Eq.\ (\ref{phirho}) is wrong. Steps [4.(a)] and [5.]\ are used elsewhere in quantum mechanics. For example, constructing the state of a sub-ensemble from probabilities conditioned on a future measurement via QT is standard procedure in delayed-choice entanglement swapping (D-CES) experiments \cite{peres2,ma,megedish}. Therefore it must be the steps [1.]\ and [2.]\ (and applying them in steps [3.]\ and [4.(b)]) which are in error. In other words the error was due to assigning a state to a qubit according to [1.] above in the first place, i.e.\ adopting the QSI is not viable. Our argument can easily be generalised to higher dimensional Hilbert spaces and so our most important result is that quantum systems cannot be assigned states. 

Pre-entanglement as presently defined also occurs in D-CES experiments \cite{peres2,ma,megedish}. In both cases, the initial states of each member of a pair is in a maximally-mixed state (a proper mixture according to the QSI for the present case and an improper mixture for D-CES). The difference is that in the present experiment but not D-CES (i) the same pair of qubits is involved in the both the pre-and post-entanglement, (ii) each member of a pair can be prepared in a way chosen by the experimenter and (iii) the choice of preparation can be made independently (e.g.\ at space like separation). It is these differences which allow us to reach the above conclusion. 

The state of an ensemble or sub-ensemble is determined by the criteria used to select the ensemble. An ensemble or sub-ensemble selected on the basis of the preparation in the present case will be separable. The sub-ensemble $i \in \{\Phi^+\}$ selected from the full, maximally mixed ensemble on the basis of the subsequent entangling measurement outcome will be entangled. That the state of a sub-ensemble depends on the selection criteria used to form the sub-ensemble is familiar from the (improper) ensembles involved in post-entanglement.  For example, if measurements of $\hat{\sigma}_{i}$ and $\hat{\sigma}_{j}$ are performed on one a pair of entangled qubits, the other qubit can be selected into four quite different sub-ensembles in the states $\ket{i^{\pm }}$ and $\ket{ j^{\pm }}$. Our result is that this procedure of ensemble selection and consequent state assignment applies in other contexts, such as the one considered in Fig.~1. Some other interpretations of quantum mechanics \cite{fuchs,peresstate} which do not assign states to individual quantum systems are also consistent with the present result.

Our result impacts on the recent theorem of Pusey, Barrett and Rudolph (PBR) \cite{pbr}. In terms of the ontological model concept of Harrigan and Spekkens \cite{harrigan}, PBR assume that the preparation of a quantum state $\ket{\psi_i}$ according to the QSI results in the quantum system also acquiring an \textit{ontic} state $\lambda$ sampled from a probability distribution $\mu_i(\lambda)$. A second assumption of preparation independence (PI) is that if two or more systems are prepared independently, the ontic state distribution is the product $\mu_1(\lambda)\mu_2(\lambda)\ldots$. The important and surprising result \cite{pbr} of those assumptions is that the supports of the $\mu_i(\lambda)$ are disjoint. Consequently, $\mu_i(\lambda)$ uniquely determines $\ket{\psi_i}$ and so $\ket{\psi_i}$ itself must be regarded as ontic.

The PBR result requires an entangling measurement and hinges on which of the prepared states lead to which of the entangled outcomes. Our results conflict with two of the PBR assumptions. Firstly, we claim that only sub-ensembles have states, not individual quantum systems. One could perhaps attempt to assign ontic states to ensembles but then the PBR reasoning would mean that the probability distributions of the ensembles involved in the theorem had disjoint support and therefore the \textit{ensemble} states should be regarded as ontic. Since the state of an ensemble could not be regarded as ontic, the combination of (i) only ensembles have states and (ii) states are ontic would mean that there are no ontological models of quantum mechanics of the form proposed by Harrigan and Spekkens \cite{harrigan}.

However that consequence does not apply because our results also call into question the assumption of PI required for the PBR theorem. The PBR protocol requires $n$ independent sources which can be randomly prepared in either of the two non-orthogonal states $\ket{\psi_0}$ and $\ket{\psi_1}$.  To relate that to the present result, let independent sources prepare maximally-mixed states by producing equal numbers of states in the basis $(\psi_0,\psi^{\perp}_0)$ or in the basis $(\psi_1,\psi^{\perp}_1)$, where $\langle \psi^{\perp}_i|\psi_i \rangle=0$. As above, it is possible to identify sub-ensembles that are in entangled states by conditioning on the outcome of the entangling measurement required for the PBR theorem. Therefore, each of the $n$-tuple input states is a member of an entangled sub-ensemble so PI could not apply to any $n$-tuple. Nevertheless, the results required for the PBR theorem could be obtained by simply eliminating the runs in which $\psi^{\perp}_0$ or $\psi^{\perp}_1$ were involved. Eliminating half of the runs in that way would not alter the fact that the remaining $n$-tuples could have been in entangled sub-ensembles which implies PI could not apply to them.

Note that the failure of PI occurring here for the pre-entangled sub-ensembles is at the quantum level not just the ontic level required for the PBR theorem. Since the pre- and post-entanglement states are the same, the failure of PI must be at the non-signalling level. Emerson \cite{emerson} and Mansfield \cite{mansfield} have shown that non-classical correlations at the non-signalling level are sufficient to prevent the derivation of the PBR theorem. Unlike other investigations of the failure of PI \cite{lewis,emerson,mansfield}, preparation dependence is here confined to systems which are subsequently entangled.  If PI is not assumed \cite{lewis} or is weakened \cite{mansfield,emerson}, (admittedly contrived) models in which $\psi$ is not ontic are possible \cite{lewis,emerson}. Therefore some non-trivial assumption is required \cite{lewis,ballentine3} if it is to be shown that quantum states are ontic. Quite different assumptions from those which lead to the PBR result are currently being actively explored (see Ref.~\onlinecite{eric} and references therein). It will be interesting to investigate those different assumptions in relation to the ESI and pre-entanglement.

Another interesting question is how pre-entanglement comes about. It seems that the pairs of quantum systems have to anticipate in some sense that they are going to be entangled in the future. The idea that measurements can have retrospective (backwards-in-time) influences has been suggested in other contexts in quantum mechanics. These include the nonlocal correlations between pairs in conventional (post-)entangled states (for a discussion, see Ref.~[\onlinecite{price}]) and as a quantum information link \cite{bennett,penrose,svetlichny} in quantum teleportation and other experiments.

Our main result is that independently prepared quantum systems can be in a (pre-)entangled state prior to an entangling measurement performed on them. If individual quantum systems possessed quantum states, their independent preparation would mean they were in a separable state and therefore could not be in an entangled state. This dilemma is avoided if quantum states are assigned only to ensembles of quantum systems. Ensembles selected in different ways can then have different quantum states from which the probabilities of results of experiments performed on the ensembles can be calculated in the usual way. In particular, selecting the ensemble on the basis of a subsequent entangling measurement leads to a ``pre-entangled" ensemble state.  All mixtures in quantum mechanics should be understood in the same way as the reduced states of the constituents of an entangled state are currently understood. That is, all mixtures in quantum mechanics are improper.

We thank E. Cavalcanti for a very helpful discussion and further comments. M.\ F.\ is supported by the John Templeton Foundation through the project \textit{New Agendas for the Study of Time}.

\end{document}